# ELECTROCHEMICAL DEGRADATION OF METHYLENE BLUE USING Ce(IV) IONIC MEDIATOR IN THE PRESENCE OF Ag(I) ION CATALYST FOR ENVIRONMENTAL REMEDIATION

Henry Setiyanto[1,2*], Feni Mustika Sari[1], Muhammad Yudhistira Azis[1], Ria Sri Rahayu[1], Amminudin Sulaeman[1], Muhammad Ali Zulfikar[1], Diah Ratnaningrum[3] & Vienna Saraswaty[3*]

[1] *Analytical Chemistry Research Group, Bandung Institute of Technology, Bandung, Indonesia*

[2] *Center for Defense and Security Research, Bandung Institute of Technology, Bandung, Indonesia*

[3] *Research Unit for Clean Technology, Indonesian Institute of Sciences, Bandung, Indonesia*

[*]*Corresponding author: henry@chem.itb.ac.id, vsaraswaty@gmail.com*

**Abstract**

Methylene blue (MB) is often used in textile industries and is actively present in the wastewater runs-off. Recently, mediated electrochemical oxidation (MEO) offers a fast, reliable and promising results for environmental remediation. Thus, we aimed to evaluate the electro-degradation potential of MB by MEO using Ce(IV) ionic mediator. Furthermore, we also observed the influence of addition Ag(I) ion catalyst in MEO for degradation of MB. The electro-degradation of MB was evaluated by cyclic voltammetry technique and was confirmed by UV-Vis spectrophotometry, high performance liquid chromatography (HPLC) analysis and back-titration analysis. The results showed that in the absence of Ag(I) ion catalyst, about 89 % of MB was decolorized within 30 minutes. When 2 mM of Ag(I) ion catalyst was applied, the electro-degradation of MB was increased to maximum value of 100%. The UV-Vis spectrum confirmed the electro-degradation of MB as suggested by decreased maximum absorbance value at λ 668 nm from 2.125 to 0.059. The HPLC analysis showed the formation of five new peaks at retention time of 1.331, 1.495, 1.757, 1.908 and 2.017 minutes, confirming the electro-degradation of MB. The back-titration analysis showed about 52.9% of $CO_2$ was produced during electro-degradation of MB by MEO. More importantly, more than 97% of Ce(IV) ionic mediator were recovered in our investigation. Our results reveal the potential of MEO using Ce(IV) ionic mediator to improve the wastewater runs-off quality from textile as well as other industries containing methylene blue.

**Introduction**

Dye pollutants produced by textile industries becoming important issue in environmental contamination. About 36.000 tons/year of methylene blue (MB) (See Fig. 1), a basic and cationic dye, are heavily used in textile industries (Murali and Uma, 2016; Yaseen and Scholz, 2019). It was estimated that approximately 10 to 20% of MB were released in water without treatment (Ehrampous et al., 2011; Güyer et al. 2016; Reza et al. 2016; Alam and Hossain et al. 2018), resulting in water pollution (Forgacs et al. 2004; Murali and Uma, 2016; Gita et al. 2017; Hassaan and Nemr, 2017; Lellis et al. 2019). In addition, in the terms of health, direct exposure of MB on skin, eyes, as well as respiratory and digestion tract was reported causing irritation, nausea as well as diarhea (Hamdaoui and Chiha, 2006). Thereby making the remove of MB from wastewater runs-off become essential to prevent environment and health problems. Several approaches, including physico-chemical treatment (membrane filtration, coagulation, adsorption and flocculation), biological treatment, and chemical oxidation have been found to effectively remove MB from wastewater (Panizza and Cerisola, 2008; Setiyanto et al. 2016; Sivagami et al. 2018; Cuerda-Correa et al. 2020). However, these conventional treatments are time consuming, expensive and require complex procedures. Additionally, some of them causing secondary environmental problem, for example by using microbial waste treatment, it may result on bad odor and insect, disturbing ecosystem, producing foams as well as reducing oxygen transfer. And some physically waste treatment such as filtration and adsorption, although vulnerable for treatment of MB, however, those method may also result in sludge formation and requires a regular generation (Bache et al. 1991; Bousher et al. 1997). For the aforementioned reasons, a simple, rapid and cost-effective method for removing MB from wastewater runs-off need to be developed.

*Figure 1*

Many scientific investigations have reported the emerging potentials of electrochemical method for solving analysis problems, studying reaction mechanisms as well as performing organic compounds degradation (Setiyanto et al., 2011; Azab et al., 2019; Kassa and Amare, 2019; Setiyanto et al. 2020). In addition, electrochemical method offers a rapid, an efficient and a low-cost degradation of organic compounds (Matheswaran et al., 2008; Möhle et al. 2018; Syaifullah et al. 2018; Setiyanto et al., 2018). A study reported that mediated electrochemical oxidation (MEO) is one of a suitable-processes for destruction of organic pollutant. By applying MEO the degradation of organic pollutants is processed indirectly through a distinctive mediator and membrane, hence metal ions emitted by ionic mediators are no longer produced and secondary environmental problem can be avoided (Matheswaran et al. 2008).

Several potential ionic mediators for organic compound degradation by MEO are Co(III)/Co(II), Ce(IV)/Ce(III) and Ag(II)/Ag(I). Those ionic mediators are reported act as oxidants and degrade organic compounds into carbon dioxide and water in an electrolyte solution. Ce(IV), as an ionic mediator, has advantages, that are easy to recover, reuse and highly soluble in electrolyte solutions (Paulenova et al. 2002; Xie et al., 2011; Palanisami et al. 2015).

Some reports also showed that the application of metal ion catalysts can accelerate the degradation of organic compounds by MEO, thereby producing a better efficiency process (Panizza and Cerisola, 2009; Zhou et al. 2013; Shestakova and Silanpää, 2017). Due to electroactive nature of MB, it is possible applying MEO for degradation of MB using Ce(IV) ionic mediator. Since, no report found about the potential of Ce(IV) as ionic mediator as well as the influence of Ag(I) ion catalyst for electrochemical degradation of MB, here, we present the electro-degradation potential of MB by MEO using Ce(IV) ionic mediator in the presence of Ag(I) ion catalyst. The electrochemical degradation of MB was confirmed using UV-Vis spectrophotometry, high performance liquid chromatography (HPLC) analysis and back titration analysis. In addition, the percentage recovery of Ce(IV) ionic mediator was also observed.

## Materials and Methods

### Apparatus

The electrochemical degradation of MB was conducted by using Potensiostat/Galvanostat eDAQ 410 with Pt wire as the cathode and anode. The electrochemical degradation product of MB was evaluated by voltametric analysis using BASi Epsilon Electrochemical Analyzer. In the voltammetry experiments used three electrodes: carbon paste electrode (CPE) as the working electrode, Ag/AgCl as the reference electrode, and Pt wire as the auxiliary electrode. The product of the electrochemical degradation of MB was further evaluated using UV-Vis spectrophotometer and high-performance liquid chromatography (HPLC) Agilent Technology 1260 instrument.

### Chemicals

Methylene blue (MB) powder and cerium sulphate octahydrate (>99.0%) were purchased from Sigma Aldrich with pro analysis grade. Silver nitrate, sulfuric acid, hydrochloric acid, nitric acid, potassium chloride, acetonitrile, glacial acetic acid and formic acid were obtained from E. Merck (Darmstadt, Germany) with pro analysis grade. Whereas platinum wire, silver wire and copper wire were obtained from PT Aneka Tambang Tbk.

## Methods

### *Preparation of the working and reference electrode*

For voltametric investigation, a carbon paste electrode (CPE) as a working electrode was prepared by heating the mixture of graphite powder and liquid paraffin (at ratio of 7:3) at 80 °C in a beaker glass. Subsequently, the mixture of CPE was inserted to a cylindrical electrode holder and cooled to room temperature. For the electrical contact, the CPE was connected to a copper wire and the surface of CPE was then polished by alumina slurry onto a polishing pad and smoothed by rubbing it on a clean paper sheet. Whereas the home-made reference electrode Ag/AgCl was prepared by electrolyzing the Ag wire in a 0.1 M KCl solution at 2 V. The home-made reference electrode Ag/AgCl was characterized by cyclic voltammetry and compared with a commercial Ag/AgCl electrode namely bioanalytical standard (BAS) prior the electrochemical degradation of MB.

### *Optimization of supporting electrolyte for MEO*

Optimization of supporting electrolytes was conducted by testing the cyclic voltammetry behaviour of 0.01 M Ce(III) solution in acidic medium that are $H_2SO_4$, $HNO_3$ and $HCl$. The supporting electrolyte which gives the highest difference in anodic peak current ($\Delta I_{pa}$) was selected for further evaluation.

### *Sample preparation*

The solution of MB was prepared by dissolving 25 mg of MB powder in 16.7 mL of 1.8 M sulphuric acid. Subsequently, the solution was added with 83.3 mL of distilled water. The MB solution in sulphuric acid was then used for electrochemical degradation by MEO and other analysis.

### *Electrochemical degradation of MB*

The electrochemical degradation of MB was conducted by using Potensiostat/Galvanostat eDAQ 410 with Pt wire as the cathode and anode. MB solution at 250 ppm was used in this study. The degradation process was conducted using Ce(IV) as the ionic mediator and Ag(I) as the ionic catalyst. The percentage degradation of MB was calculated using the following formula

$$\%Degradation = \frac{I_0 - I_t}{I_0} x\ 100\%$$

Where $I_o$ is the current of initial MB oxidation peak (before electrochemical degradation) and $I_t$ is the current of MB oxidation peak after the electrochemical degradation process.

### *Calculation of Ce(III)/Ce(IV) recovery*

The percentage recovery of Ce(III)/Ce(IV) was calculated using the following formula:

$$\%\ Recovery = \frac{\Delta Ipc}{\Delta Ipa} x 100\ \%$$

Where $\Delta Ipc$ is cathodic peak current at E 1.17 V and $\Delta Ipa$ is anodic peak current at 1.25 V.

### *UV-Vis spectrophotometry analysis*

To confirm the electro-degradation of MB was successful, we recorded the absorbance of MB solution (before and after electro-degradation) at a wavelength of 200 to 800 nm by using UV-Vis Spectrophotometer (Hitachi U2800, Japan).

## Back titration analysis

The back titration analysis (the Ba(OH)$_2$ titration method was used to confirm and calculate the formation of CO$_2$ after electrochemical degradation of MB. The amount of Ba(OH)$_2$ is assumed equivalent with CO$_2$ according the following reaction :

$$Ba(OH)_2\ (aq) + CO_2\ (g) \rightarrow BaCO_3\ (s) + H_2O.$$

The procedure for determination of CO$_2$ is as follows: the CO$_2$ (g) formed during electrochemical degradation of MB was fed into an erlenmeyer flask containing Ba(OH)$_2$ solution. After electrochemical degradation of MB completed, the erlenmeyer flask of Ba(OH)$_2$ solution containing CO$_2$ was then titrated with HCl solution. The amount of Ba(OH)$_2$ remained was assumed as the amount Ba(OH)$_2$ of which did not react with the CO$_2$ formed from electrochemical degradation of MB. The percentage of CO$_2$ formed was calculated according to the following formula :

$$\%\ CO_2 = \frac{[\text{mmol Ba(OH)2 total} - \text{mmol Ba(OH)2 titration}]}{mmol\ (Ba(OH)2\ total)} \times 100\%$$

## HPLC Analysis

HPLC analysis was also performed to confirm and to detect the formation of new compound as the result of electrochemical degradation of MB. The HPLC analysis was performed by using HPLC Agilent Technology – 1260 instrument with C18 Column. About 20 µL of MB was injected into HPLC instrument. The mixture of 0.1 M ammonium acetate pH 5.3 (A) and acetonitrile (B) was generated as a mobile phase. The elution gradient was adjusted from 5:95 to 95:5 with flow rate of 1 mL/min for 15 min. The degradation products of MB were confirmed by comparing the HPLC chromatogram of MB before and after electro-degradation of MB by MEO.

## RESULTS AND DISCUSSION

### *Characterization of reference electrode Ag/AgCl*

*Figure 2*

Prior to electro-degradation of MB, the home-made reference electrode used should have similar quality to commercial standard reference electrode. In order to confirm this, the cyclic voltammogram profiles of the home-made Ag/AgCl reference electrode was compared with BAS in 0.1 M KCl solution with a scan rate of 100 mV/s. As shown in Fig. 2A, a similar quasi-reversible couple appears in the cyclic voltammogram profiles of home-made reference electrode Ag/AgCl and BAS. Those results were emphasized in the calculation of the ratio of cathodic and anodic peak current ($I_{reverse}/I_{forward}$) values of home-made Ag/AgCl and BAS that were 0.8500 µA and 0.8530 µA, respectively. The peak currents and oxidation/reduction potential data of home-made Ag/AgCl and BAS in Table 1 also showed similar results and has no significant difference. Here, we confirm that the home-made prepared Ag/AgCl and BAS are completely preserved, showing similar quality.

*Table 1*

### *The best supporting electrolyte for Ce(III)/Ce(IV) mediator ion system*

Prior to electro-degradation of MB, we also evaluate the best supporting electrolyte which reduce the migration of current due to electrostatic movement. In this experiment we observed the peak current of 0.01 M Ce(III) in various supporting electrolytes that are $H_2SO_4$, HCl and $HNO_3$ at 0.2 M. As depicted in Figure 2B., the cyclic voltammogram profiles of Ce(III) solution in various supporting electrolyte are different. In addition, the calculation of difference in anodic peak current (ΔIpa) values of Ce(III) in $H_2SO_4$, HCl and $HNO_3$ are 68.72 µA, 42.00 µA and 31.42 µA, respectively. Accordingly, $H_2SO_4$ is the best supporting electrolytes for Ce(IV) ionic mediator in MEO as suggested by the highest ΔIpa value. We also noticed that the addition of Ce(III) in $H_2SO_4$ solution in electrochemical reaction resulted in two peaks in cyclic voltammogram profiles as suggested by peak at +1.25 V and +1.17 V (see Figure 2C, red line). The peak at 1.25 V refers to oxidation peak. Whereas the peak at +1.17 V refers to reduction peak. Theoretically, the reduction potential of Ce(IV)/Ce(III) is +1.24 V in accordance with Ag/AgCl electrode in 1.0 M $H_2SO_4$ solution. Hence, our results were in the line with previous study which reported that $H_2SO_4$ is used for redox reaction of Ce(IV)/Ce(III) (Ren and Wei, 2011).

### *Cyclic voltammetry behaviour of MB*

We also observed the cyclic voltammetry behaviour of MB. This analysis is important to determine the peak oxidation potential of MB. As shown in Figure 2D, the peak oxidation of MB presents at E of +0.226 V. This oxidation potential was then used as the reference for calculation of percentage degradation of MB in the presence of Ce(IV) ionic mediator. As also depicted in Figure 2D, no reduction peak was observed in Figure 2D, showing irreversible reaction of MB.

### *The optimum supporting electrolyte concentration*

Prior to the electrochemical degradation of MB, we optimized the concentration of supporting electrolyte. The concentration of supporting electrolyte plays important role in electrochemical reaction, particularly in resulting current (Ip) value as well as treatment cost. Theoretically, the higher concentration of supporting electrolytes used, the higher current was observed and the higher cost is required. We observed the effect of $H_2SO_4$ concentration at 0.1, 0.2, 0.3, 0.35, 0.4, 0.45 and 0.5 M. As shown in Fig.3, at concentration of 0.1 to 0.3 M, the Ip values tend to increase. When the concentration of $H_2SO_4$ was raised from 0.35 to 0.5 M, the curve tends to be flat. Since the curve of Ipa values of $H_2SO_4$ starting to be flat at 0.3 M, we therefore applied 0.3 M $H_2SO_4$ for further evaluation.

### *Electrochemical degradation of MB*

The potential electro-degradation of MB using Ce(IV) ionic mediator was observed as function of working potential at 0.1 to 10.5 V. In this experiment, the Ce(IV) ionic mediator, the active species, takes place indirectly in electrochemical oxidation of MB. In acidic solution Ce(III) ion is oxidized become Ce(IV) ion. Ce(IV) ion is a strong oxidant species. In redox system, when Ce(IV) is reduced become Ce(III), MB will be oxidized electrochemically, generate its reducing form. We also compare the electro-degradation result of MB in the absence of Ce(IV) ionic mediator. It is obviously depicted in Figure 4, the percentage electro-degradation of MB using 0.01 M Ce(IV) ionic mediator is higher than that of without Ce(IV) ionic mediator. Furthermore, as also depicted in Figure 4, the percentage electro-degradation of MB tends to increase along with the higher value of working potential. This because, in electrochemical reaction, the voltage is the driving force and acts as the power of repolarization of the particle electrode. When the voltage is increased, the degree of repolarization is enhanced (Liu et al., 2012). Therefore, the electrochemical reaction of MB is improved, so the reaction speed is also accelerated, then enhancing the percentage degradation of MB. However, when the working potential was increased from 7.5 V to 10.5 V, electro-degradation of MB entered the steady-state phase. It seems by using working potential at 7.5 V the optimum electrochemical reaction reached. Therefore, we recommend applying a 7.5 V working potential in the presence of Ce(IV) ionic mediator for electro-degradation of MB.

### *The influence of Ce(III) concentration, time and the percentage recovery*

In environment remediation, chemicals concentration used and treatment time play important role in particular in cost effectiveness. The lowest concentrations of chemicals and the shortest treatment time, the lower cost is required. Therefore, we further optimized the concentration of Ce(III) and time for electro-degradation of MB by MEO at 250 ppm. As shown in Figure 5A, the higher Ce(III) concentrations, the higher degradation of MB was observed. However, when Ce(III) at 0.02 M to 0.03 M was applied, the percentage degradation of 250 ppm MB at working potential of 7.5 V tends to be flat, indicating no more MB was degraded. We also noticed that the percentage degradation of 250 ppm MB showed similar trend at function of time. As shown in Fig. 5B, at the first of 30 mins, the percentage degradation of MB increased rapidly then entering the steady state phase at 30-60 minutes with ~89% of MB was degraded. Suggesting, for electro-degradation of MB at 250 ppm by MEO using Ce(IV) ionic mediator, about 0.02 M Ce(III) and a minimum of 30 minutes treatment is required.

We also further evaluated the recovery of Ce(IV) ionic mediator, because this evaluation is important to confirm that the second pollution is not generated using the MEO method. As suggested in Fig. 2C, the oxidation and reduction peak currents of Ce(III)/Ce(IV) lie at E values of +1.25 and 1.17 V respectively. Hence, we further observed the value of Ipc and Ipa of Ce(III)/Ce(IV) and presented the values and calculated the percentage recovery of Ce(IV) ion in Table 2. As can be seen in Table 2, the percentage recovery of Ce(IV) as ionic mediator was in the ranging of 97.1 to 101.6%. Hence, we confirmed, electrodegradation of MEO did not generate second pollution. However, in this experiment, we used one pot reaction system. Therefore, we further suggest to use two pots reaction system using a membrane to separate Ce(IV) ion after processing.

### *The presence of Ag(I) ion catalyst enhanced electrochemical degradation of MB*

Ag(I) ion may act as ionic mediator as well as ion catalyst. Studies have shown that the presence of ion catalyst in mediated electrochemical oxidation can accelerate the degradation of organic compounds then enhancing its degradation potential. We selected Ag(I) as ion catalyst because it has a higher E° standard value than Ce(IV) ion. In this experiment, we observed the influence of Ag(I) toward MB degradation using MEO in the presence of Ce(IV) ionic mediator at 1 to 2.5 mM. As depicted in Fig. 5C, the percentage degradation of MB increases along with the increase of Ag(I) concentration. More importantly, when 2 mM Ag(I) was applied, the degradation of 250 ppm MB reached 100%. It should be noted that the standard reduction potential (E°) value of Ag(II)/Ag(I) and Ce(IV)/Ce(III) in aqueous solution are 1.98 V and 1.61 V respectively (Bard et al. 1985; Matheswaran et al. 2007). The higher standard E° value of Ag(II)/Ag(I), the earlier Ag(II) will be reduced become Ag(I), then augmenting the efficiency of Ce(III) conversion. Hence, undoubtly the electro-degradation of MB improved. Thus, in this experiment, Ag(I) has shown its function as catalyst which enhancing the electrodegradation of MB by MEO.

*Table 2*

However, when Ag(I) was applied at of 2.5 mM, a decreasing in electrodegradation of MB was observed. It seems that the saturation of the Ag(I) solution in the mixture corresponding to the lower result of MB degradation. In this experiment, the electrochemical degradation of MB was carried out in a batch system containing $H_2SO_4$ solution and Ce(III)/Ce(IV) ionic mediator. The percentage recovery of Ce(III)/Ce(IV) was in the ranging of 97.1 to 101.6% as suggested by comparative Δ Ipc and Ipa value (See Table 2). Therefore, the excess Ag(I) probable reacted with $H_2SO_4$ solution resulting in $Ag_2SO_4$. A study has shown that the solubility of $Ag_2SO_4$ starts to decrease in $H_2SO_4$ solution at 1.5 mM, confirming the formation of $Ag_2SO_4$ (s) (Lietzk & Stoughton, (1957). This also emphasized with the small Ksp values $Ag_2SO_4$ of $1.5 \times 10^{-5}$. For the aforementioned reason, the increase excess concentration of Ag(I) led to decrement MB degradation.

***Electrochemical degradation of MB by MEO confirmed by UV-Vis Spectrophotometer, HPLC analysis and Ba(OH)$_2$ titration***

UV-Vis spectrum and HPLC analysis obviously prove that the potential electro-degradation of MB. As depicted in Figure 5D the UV-Vis absorption bands of MB before electro-degradation shows maximum absorption peak at λ 668 nm with absorbance value of 2.125. After electro-degradation, the absorption peak at λ 668 nm reached to the minimum value of 0.059, corresponding to degradation or bleaching of MB. Whereas, the HPLC analysis obviously showed the formation of five new peaks at retention time of 1.33, 1.49, 1.75, 1.91 and 2.01 min after electro-degradation of MB (See Figure 6). According to the propose mechanism of MB degradation from previous report, by dissolving MB in water, the Cl$^-$ is separated first. Subsequently, the N-CH$_3$ bond is broken then the -CH$_3$ is oxidized into HCHO or HCOOH. Further, the C-S and C-N bond on the central heterocycle of MB are broken by free radical attack to produce 2,5-diaminobenzenesulfonic acid and 4-aminocatechol (Huang et al., 2010; Song et al., 2017). Those intermediates are oxidized to a single ring structure and finally to $CO_2$ and $H_2O$ (Teng et al., 2020). Thus, the peaks present in HPLC chromatogram of MB after electro-degradation undoubtedly relate with intermediates. The schematic diagram of the electro-oxidation mechanism of MB degradation and its intermediate products formation can be seen in Figure 7.

*Figure 7*

We also confirmed the formation and calculated the amount of $CO_2$ produced during electro-degradation of MB. This analysis is important, to prove the complete oxidation reaction of MB in producing $CO_2$. Below, we propose the reaction of MB degradation as suggested by Chung et al., 2000 and Balaji et al., 2007:

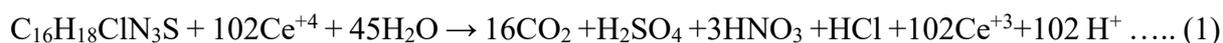

$C_{16}H_{18}ClN_3S + 102Ce^{+4} + 45H_2O \rightarrow 16CO_2 + H_2SO_4 + 3HNO_3 + HCl + 102Ce^{+3} + 102\ H^+$ ….. (1)

Here, we confirm the production of $CO_2$ as suggested by the precipitation of white sendiment which resulted from the reaction of $Ba(OH)_2$ with $CO_2$. This result was also strengthened by the calculation of $CO_2$ formed during electro-degradation of MB. It shows that MB produced about 52.88% of $CO_2$. Results from our study prove that electro-degradation of MB resulted in $CO_2$ as well as other intermediate molecules.

Electrochemical degradation of MB has been previously studied using chlorine on a TiRuO2 oxide anode. This study also demonstrated a success of MB degradation. However, the electro-degradation of MB was achieved due to the high bleaching properties of active chlorine (Panizza et al. 2007). Chlorine is highly active gas and toxic. The environment toxicity of chlorine is low. However, the exposure of chlorine gas for a short period of time into human respiratory system is dangerous, which may affect to lung irritation, resulting cough as well as chest pain. Therefore, our method for degradation of MB is more promising and environmentally friendly.

## Conclusion

This study demonstrated the potential electrochemical degradation of MB by MEO using Ce(IV) ionic mediator. The 100 % of MB degradation at 250 ppm MB was achieved within 30 minutes when 2mM Ag(I) ion catalyst was applied. More importantly, the percentage recovery of Ce(IV) ion mediator in this study is higher than 97%, thereby secondary environment pollution can be avoided. In summary, MB degradation by MEO using Ce(IV) ionic mediator in the presence of Ag(I) ion catalyst is a promising and more environmentally friendly method.

## Acknowledgements

This research was funded by the Analytical Chemistry Research Group, Faculty of Mathematics and Natural Sciences, Institut Teknologi Bandung, Indonesia. Acknowledgments are also directed to the Penelitian Dasar Unggulan Perguruan Tinggi (PDUPT) Research Grant 2019-2020 with Contract No.: 1170e/I1.C.01/PL/2019 Kementerian Riset, Teknologi dan Pendidikan Tinggi Republik Indonesia. Authors also thank to e-Layanan Sains (ELSA) LIPI for research facility.

**Conflicts of interest**

The authors declare that they have no conflict of interest.

**Author contributions**

**HS & VS**: concept, design, data collection, data analysis, manuscript writing, **FMS**: Data collection and analysis, **DR, MYA, RSR, AS, and MAZ**: Data analysis.

LIST OF FIGURES

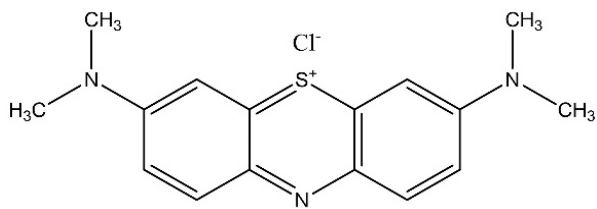

Figure 1. Chemical structure of methylene blue (MB)

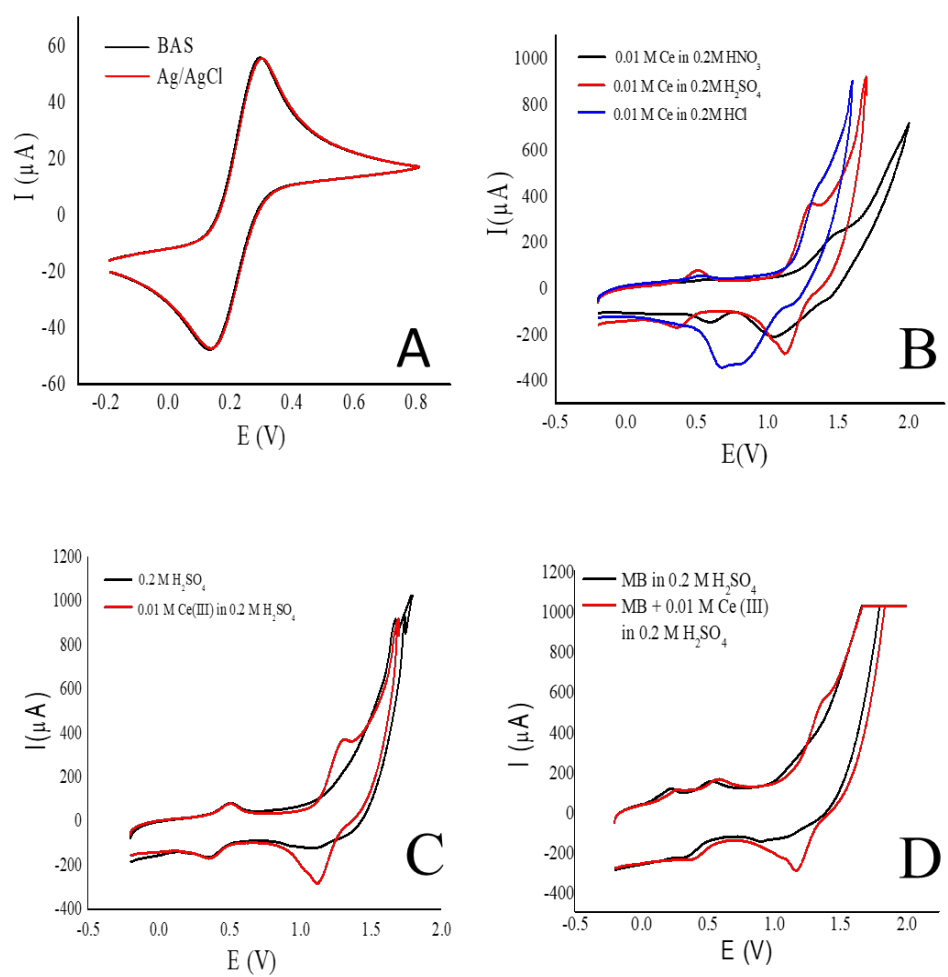

Figure 2. The cyclic voltammogram profiles of the home-made Ag/AgCl and BAS in 0.1 M KCl (A); 0.01 M Ce(III) in various supporting electrolyte solutions of 0.2 M $H_2SO_4$, $HNO_3$ and HCl (B); 0.2 M $H_2SO_4$ and 0.01 M Ce(III) in 0.2 M $H_2SO_4$ (C); MB and MB+ 0.01 M Ce(III) in 0.2 M $H_2SO_4$ (D) at a scan rate of 100 mV.sec$^{-1}$.

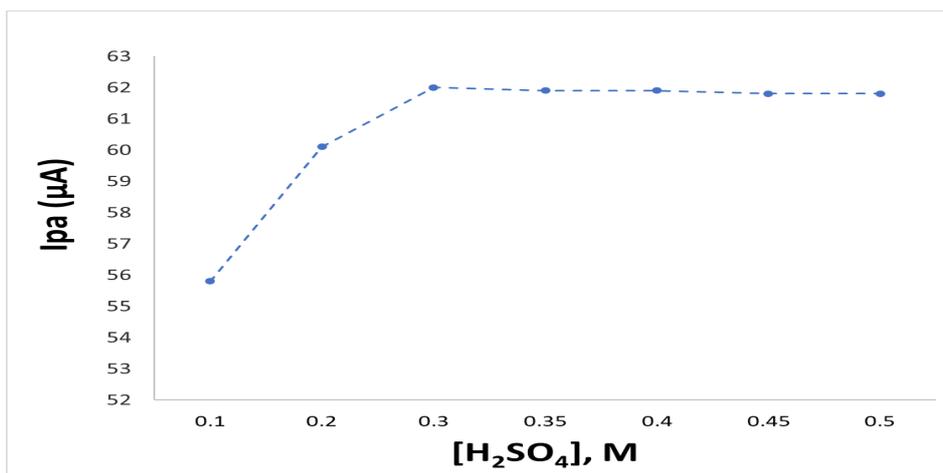

Figure 3. The Ipa value of Ce(III) as function of H₂SO₄ concentrations

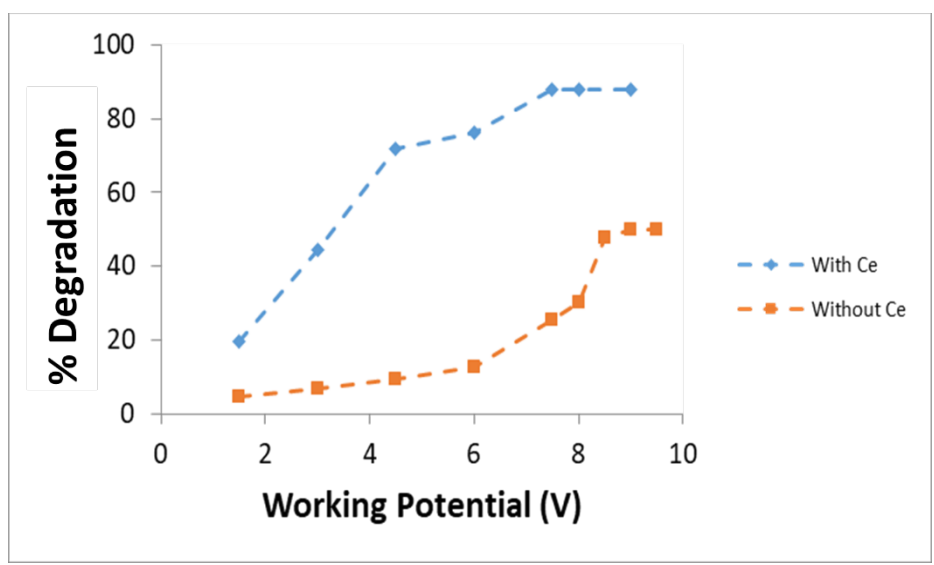

Figure 4. The percentage degradation of MB as a function of working potential (V) in the presence and absence of Ce(IV)/Ce(III) ionic mediator.

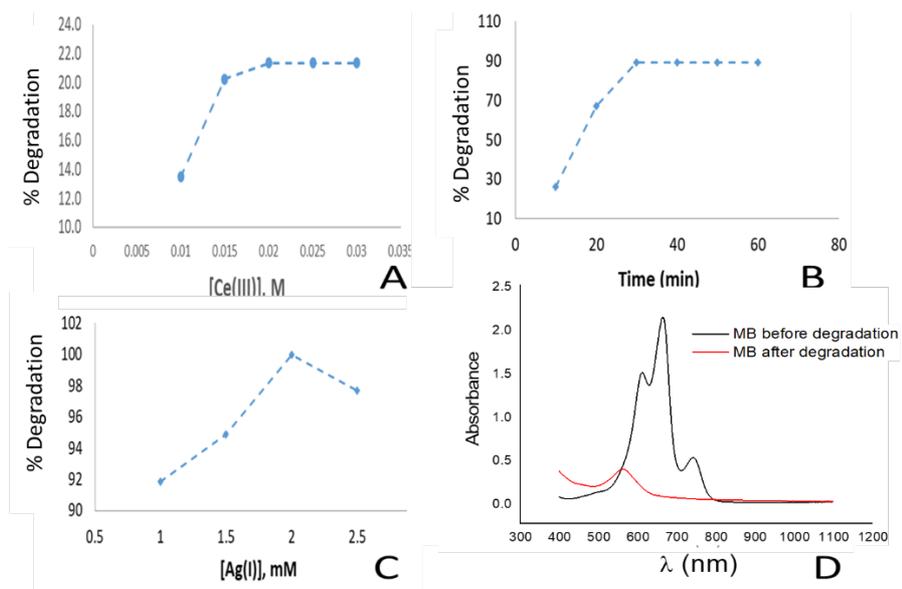

Figure 5. The percentage degradation of MB as function of Ce(III) concentration (A); Time (B) and Ag(I) concentration (C) in acidic medium $H_2SO_4$. And UV-Vis spectrum of MB before and after electro-degradation (D).

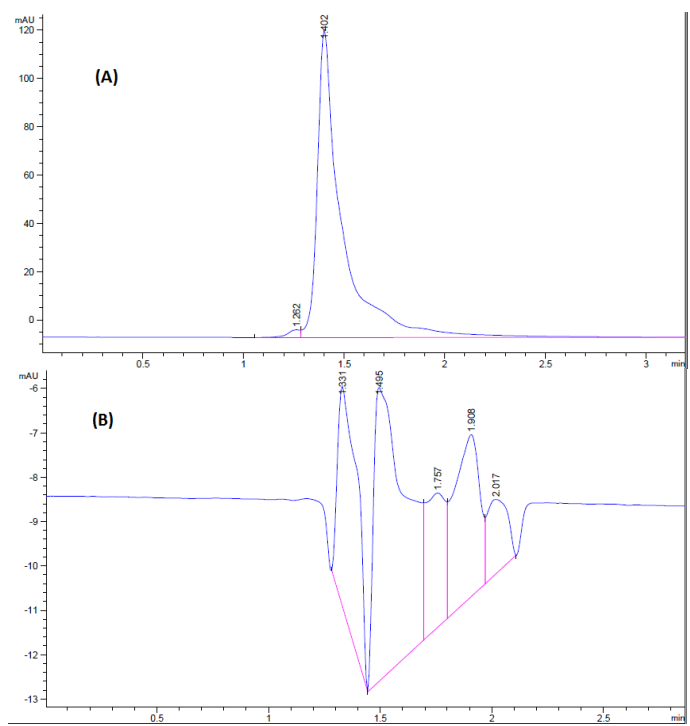

Figure 6. HPLC chromatogram of MB before (A) and after (B) electro-degradation by MEO

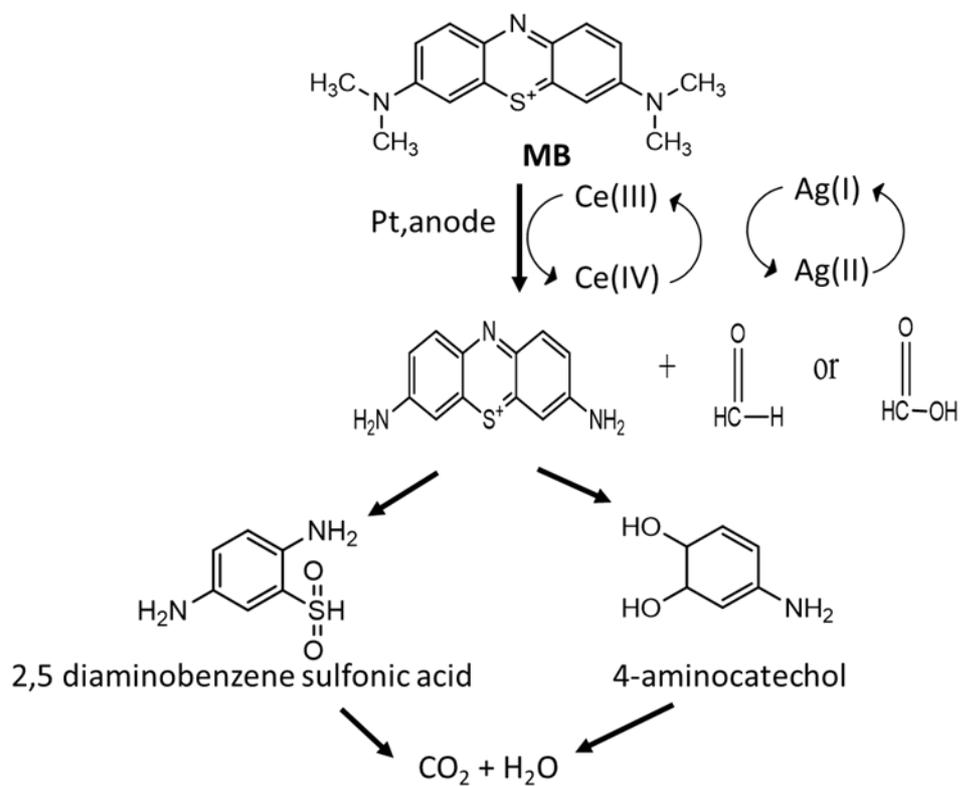

Figure 7. The schematic diagram of the electro-oxidation mechanism of MB degradation and its intermediate products formation

LIST OF TABLES

**Table 1.** The peak currents and oxidation/reduction potential of home-made Ag/AgCl and BAS

| Electrode | Ipa (µA) | Epa (V) | Ipc (µA) | Epc (V) |
|---|---|---|---|---|
| BAS | 55.78 | 0.2880 | -47.62 | 0.1240 |
| Home-made Ag/AgCl | 55.43 | 0.2940 | -47.17 | 0.1260 |

Abbreviation : I = current value, E = potential value, pa = anodic peak, pc = cathodic peak

**Table 2.** The oxidation/reduction peak currents of Ce(III)/Ce(IV) and % Recovery

| Potential Peak (Ep) | ΔI (µA) | | | % Recovery | | |
|---|---|---|---|---|---|---|
| | I | II | III | I | II | III |
| Epc (1.17 V) | - 31.5 | - 32.5 | - 30.6 | 101.6 | 101.5 | 97.1 |
| Epa (1.25 V) | + 31.0 | + 32.0 | +31.5 | | | |

Abbreviation : pa = anodic peak, pc = cathodic peak